\documentclass{article}

\usepackage{arxiv}

\usepackage[utf8]{inputenc} 
\usepackage[T1]{fontenc}    
\usepackage{hyperref}       
\usepackage{url}            
\usepackage{booktabs}       
\usepackage{amsfonts}       
\usepackage{nicefrac}       
\usepackage{microtype}      
\usepackage[numbers]{natbib}
\usepackage{graphicx}
\usepackage{amsmath}
\usepackage{siunitx}

\title{Modelling of thermosensitive stereoregular polymers within Martini Coarse-Grained force-field: poly(N-isopropylacrylamide) as a benchmark case}

\author{
  Alexander D.~Muratov \\
  N.N.Semenov Federal Research Center of Chemical Physics of Russian Academy of Sciences \\
  ul. Kosygina $4$ \\
  $119991$ Moscow, Russian Federation \\
  \texttt{ad.muratov@physics.msu.ru} \\
   \And
  Anastasia A.~Markina \\
  N.N.Semenov Federal Research Center of Chemical Physics of Russian Academy of Sciences \\
  ul. Kosygina $4$ \\
  $119991$ Moscow, Russian Federation \\
  \AND
  Dmitry V. Pergushov\thanks{Department of Chemistry, M.V.Lomonosov Moscow State University, Leninskie Gory $1/3$, $119991$ Moscow, Russian Federation.} \\
  N.N.Semenov Federal Research Center of Chemical Physics of Russian Academy of Sciences \\
  ul. Kosygina $4$ \\
  $119991$ Moscow, Russian Federation \\
  \And
  Vladik A. Avetisov \\
  N.N.Semenov Federal Research Center of Chemical Physics of Russian Academy of Sciences \\
  ul. Kosygina $4$ \\
  $119991$ Moscow, Russian Federation \\ \\
}

\begin{document}
\maketitle

\begin{abstract}
In this article we introduce the MARTINI model of a widely used thermoresponsive polymer, Poly(N-isopropylacrylamide). Importantly, our model takes into account polymer's stereoregularity (i.e. tacticity). We highlight the effect of changing the polarity of coarse-grained particles comprising the polymer on its ability to exhibit a temperature-dependent coil-to-globule transition. Our approach allows to study the behavior of thermoresponsive polymers at large time and length scales.
\end{abstract}


\section{Introduction}
Poly(N-isopropylacrylamide) (PNIPA) is a well-known thermoresponsive polymer, since it experiences a fine defined
coil-to-globule transition at its lower critical solution temperature (LCST) around 305K, near body temperature\cite{convertine_facile_2004,gao_stimuli-responsive_2017,kamath_thermodynamic_2013,zhang_active_2010,pasparakis_lcst_2020}. Thus it finds many applications in nanoscience, such as drug delivery, catalysis, sensors, and in many other fields. Hydrophobic backbone chain and isopropyl groups tend to avoid contacts with water molecules while hydrophilic amide groups tend to form hydrogen bonds. Below LCST amide groups form water-polymer hydrogen bonds while above LCST these bonds are replaced with polymer-polymer hydrogen bonds\cite{sun_chain_2010}.  PNIPA has been thoroughly studied using atomistic molecular dynamics (MD) and has been modelled using PCFF\cite{deshmukh_role_2012}, OPLS\cite{alaghemandi_molecular_2012,walter_molecular_2012}, AMBER\cite{du_effects_2010}, GROMOS\cite{kamath_thermodynamic_2013} and CHARMM\cite{kamath_thermodynamic_2013}. Among other challenges, a chain length dependence was studied\cite{deshmukh_role_2012,tucker_study_2012} as well as effects of stereoregularity\cite{chiessi_influence_2016}. Also, PNIPA behaviour in salt aqueous solution\cite{du_effects_2010} or in water-methanol mixture\cite{walter_molecular_2012,hofmann_methanol-induced_2015,mukherji_relating_2016} was also investigated. However, most of these simulations fail to reproduce a reverse globule-to-coil transition that is observed experimentally with decreasing temperature. Most probably, this is due to the fact that the transition from globule to coil is remarkably slower than the direct process of the coil-to-globule transition and thus is inaccessible by atomistic MD.

Atomistic simulations are limited in time- and length-scales. Hence, one might want to use coarse-grained (CG) simulations. There are two types of CG approaches: "bottom-up", which focuses on accurate representation of the underlying atomistic structure of the compound , and "top-down", which concentrates on reproducing macroscopic properties of the compound. An example of "bottom-up" approach application for PNIPA description was reported by \citeauthor{bejagam_machine-learning_2018}; they used the data-driven machine-learning method to parametrize intra- and intermolecular CG potentials\cite{bejagam_machine-learning_2018}. However, "bottom-up" models, accurate as they might be, are usually nontransferable; that is, they require reparametrization should any change in the conditions occur. So such a temperature-independent model fulfils the task of describing the coil-to-globule transition of PNIPA in water, but when the interaction between PNIPA and any chemical compound rather than water is concerned, it would require much additional parametrization. 

On the other hand, most "top-down" approaches are computationally cheaper and easily transferable; such methods usually exploit the idea of building blocks and use simple potential forms. Pre-parametrized blocks can be used in different molecules. Several "top-down" CG models of PNIPA have been introduced so far; most notably Abbot and Stevens have implemented a hybrid approach combining the "top-down" SDK (Shinoda-DeVane-Klein\cite{shinoda_multi-property_2007}) CG force-field for non-bonded parameters with the "bottom-up" multi-centered Gaussian-based potentials for bonded parameters\cite{abbott_temperature-dependent_2015}. This temperature-dependent model correctly reproduces the LCST behaviour of PNIPA. However, the SDK force-field is not widespread and its limited to a few compounds. 

Another widespread CG force-field is Martini that parameterizes Lennard-Jones potentials to experimental thermodynamic data such as partitioning free energy between water and octanol-1\cite{marrink_coarse_2004,marrink_martini_2007} ($\Delta \Delta G_{OW}$). Its main feature is several discrete interaction levels. The Martini force-field is applicable for lipids, proteins\cite{monticelli_martini_2008,de_jong_improved_2013}, carbohydrates\cite{lopez_martini_2009}, glycolipids\cite{lopez_martini_2013}, DNA\cite{uusitalo_martini_2015} and polymers, such as poly(ethylene oxide)/poly(ethylene glycol) (PEO/PEG)\cite{lee_coarse-grained_2009, rossi_coarse-grained_2012,nawaz_coarse-graining_2014,taddese_effect_2017}, poly(styrene) (PS)\cite{rossi_coarse-graining_2011,rossi_molecular_2012}, poly(methyl methacrylate) (PMMA)\cite{uttarwar_study_2013} or poly(acrylamide) (PAM)\cite{wang_coarse-grained_2012, banerjee_coarse-grained_2018}. However, so far there was only a few attempts to introduce PNIPA in the frame of the Martini force-field. First Martini model of PNIPA lacked important structural details such as stereoregularity since amide and backbone groups were reproduced by one bead\cite{perlmutter_all-atom_2011}. Moreover, thermoresponsitive behaviour of PNIPA was out of scope of that work.

Recently, \citeauthor{perez-ramirez_coil--globule_2020} have made an attempt to override this shortcoming by introducing another Martini model for PNIPA\cite{perez-ramirez_coil--globule_2020}; this model is temperature-independent and reproduces the LCST of the polymer. Although we do not doubt the ability of the model to reproduce the LCST in this case, we express some concerns that the ideas they implement into the model make it incompatible with the version of Martini they claim they are using. First of all, we would like to notice that the authors do not use standard values of Martini non-bonded parameters. Non-bonded interactions are represented in Martini by the Lennard-Jones potential, which is defined two numeric values, $\sigma$ and $\epsilon$ (basically, these are the distance at which the potential turns to zero and its depth, respectively; more details are given below in Section \ref{M&MA}) and in the Martini version $2.2$ $\sigma$ may acquire only two possible values. However, the authors use several values of $\sigma$, which is contradicting the Martini approach. The values of $\epsilon$ reported in the article are also different from those used in Martini. For instance, \citeauthor{perez-ramirez_coil--globule_2020} indicate the following values for describing the interaction between two SC$3$ beads, representing isopropyl group: $\sigma=\SI{0.45}{\nano\metre}, \epsilon=\SI{2.2}{\kilo\joule\per\mole}$, while \citeauthor{marrink_martini_2007} use $\sigma=\SI{0.43}{\nano\metre}, \epsilon=\SI{2.625}{\kilo\joule\per\mole}$. Such a difference is quite large, it corresponds to one level of interaction. Also, the authors use the so-called B-mapping to describe a polymer chain, while it was previously shown that another type of mapping is preferable for polymers such as PNIPA\cite{rossi_coarse-graining_2011} (see Section \ref{M&MB} for more details on the mapping). These flaws in the structural mapping and parametrization encourage us to develop a more accurate model.

In this paper we introduce a model of PNIPA that takes into account its thermoresponsive properties and potential stereoregular structure. It reproduces LCST and captures the difference between meso- and racemic diads, which is used for the first time in a CG model. Our implementation is compatible both with the most widespread Martini version, $2.2$ and the latest $3.0$ version\cite{martini3_2018}. The proposed approach can be exploited for further parametrization of different thermoresponsive compounds.

\section{Materials and Methods}
\subsection{Non-bonded parametrization}
\label{M&MA}
In the Martini force-field each CG particle represents from two to five heavy atoms with adjacent hydrogens; normally, a four-to-one mapping is used. There are four main types of particles: charged (named Q), polar (P), nonpolar (N) and apolar (C). These types are divided into four or five subtypes according to their degree of polarity or hydrogen-bonding capability. In total, there are $18$ particles in Martini $2.2$ and $20$ particles in Martini $3.0$ (also, there is a special bead for water). However, there are no fully charged groups in PNIPA, which limits the available number of subtypes to $14$ in the version $2.2$ and to $15$ in the version $3.0$ since charged particles cannot be used. The interaction between uncharged particles is described only by shifted Lennard-Jones (LJ) potential in the form
\begin{equation}
\label{eq:1}
v_{ij}(r_{ij})=4\epsilon_{ij}\bigg[\bigg(\frac{\sigma_{ij}}{r_{ij}}\bigg)^{12}-\bigg(\frac{\sigma_{ij}}{r_{ij}}\bigg)^{6}\bigg],
\end{equation}
in which $\sigma_{ij}$ represents the closest distance of approach between two particles and $\epsilon_{ij}$ is the strength of their interaction\cite{marrink_martini_2007}. Distance $\sigma_{ij}$ normally has a value of $\SI{0.47}{\nano\metre}$ between four-to-one mapped beads (which are called N-beads). However, taking into account that we map three or two atoms into one bead, we have to consider special CG particles, which are called "small", or S-beads\cite{marrink_martini_2007, rossi_coarse-graining_2011}. In such case $\sigma_{ij}$ is $\SI{0.43/0.4}{\nano \metre}$ between S-beads in Martini $2.2/3.0$. The latest $3.0$ version of Martini also includes "tiny" particles, or T-beads\cite{martini3_2018}, which have $\sigma_{ij}=\SI{0.33}{\nano \metre}$ between them. It is worth noting that Martini $2.2$ also includes T-beads\cite{uusitalo_martini_2015,alessandri_pitfalls_2019}; however, these beads are strictly exclusive for rings. In Martini $2.2$ $\sigma_{ij}=\SI{0.47}{\nano \metre}$ between beads of different sizes; in Martini $3.0$ generally (with a few exceptions) $\sigma_{ij}=\SI{0.44}{\nano \metre}$ between N-beads and S-beads, $\sigma_{ij}=\SI{0.4}{\nano \metre}$ between N-beads and T-beads and $\sigma_{ij}=\SI{0.37}{\nano \metre}$ between S-beads and T-beads.

Logically, a PNIPA monomer might be represented by three beads, one representing backbone chain (BB), second - amide group (AM) and third - isopropyl group (IP)\cite{abbott_temperature-dependent_2015}. In the latest Martini $3.0$ any of these groups is represented by T-bead (BB - due to its 2-to-1 mapping, AM/IP - since they are branched). As for Martini $2.2$, its representation is not that straightforward: technically, since S-beads are reserved exclusively for ring molecules, we would have to use N-beads. Despite that, using S-beads for representation of BB groups is an ubiquitous practice\cite{rossi_coarse-graining_2011,banerjee_coarse-grained_2018}; as for other groups, we try both N- and S-beads and report the results in Section \ref{Res:C}.

Since PNIPA changes its solubility with temperature, partitioning free energies will change with temperature as well. Although experimental values of partitioning free energies for PNIPA were measured for only a small range of temperatures, we might as well use the values obtained from atomistic MD simulations\cite{kamath_thermodynamic_2013}. Thus, to represent PNIPA at different temperatures we may have to change the choice of CG particles, thus making our model temperature-dependent\cite{abbott_temperature-dependent_2015}. Parameter-dependent models are not unknown within the Martini force-field; for instance, parametrization of amino acids depends on the secondary structure of proteins, in which they are incorporated\cite{monticelli_martini_2008}. On the other hand, some previous works on coarse-graining PNIPA introduce temperature-independent models\cite{bejagam_machine-learning_2018,perez-ramirez_coil--globule_2020}.

\subsection{Bonded parametrization}
\label{M&MB}
\begin{figure*}
\centering
\includegraphics[trim={9cm 11.25cm 0 0},clip]{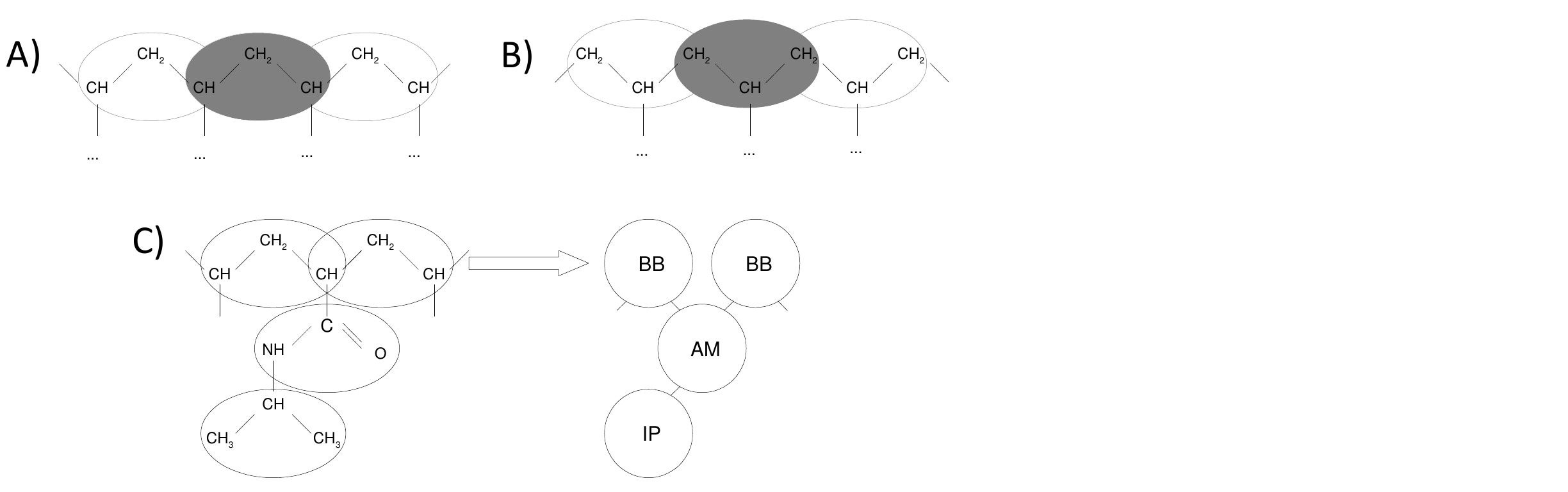}%
\caption{A-B) Examples of the mapping of the backbone polymer chain: A) the A-mapping and B) the B-mapping; C) Mapping scheme of PNIPA; here BB beads represent  the backbone chain, AM beads - amide groups, and IP beads - isopropyl groups.\label{Plot:1}}%
\end{figure*}

To describe the interactions within the molecule we introduce bonded potentials. In the Martini force-field usually two types of bonded potentials are exploited: the bond-length and the bond-angle. The bond-length interactions are reproduced in the Martini force-field as a harmonic potential:
\begin{equation}
\label{eq:2}
v_{\text{bond}}(r_{ij})=\tfrac{1}{2} k_{\text{b}}\bigg(1-\frac{r_{ij}}{r_{\text{\text{0}}}}\bigg)^{2},
\end{equation}
where $r_0$ is the equilibrium length of the bond and $k_{\text{b}}$ is the spring constant. Here it is necessary to notice that there are two possible mappings for a backbone chain: the A-mapping, in which each BB particle represents $C_{\beta}$ atom and two halves of adjacent $C_{\alpha}$ atoms(see Figure \ref{Plot:1}A)); and the B-mapping, in which each BB particle represents $C_{\alpha}$ atom and two halves of adjacent $C_{\beta}$ atoms (see Figure \ref{Plot:1}B)).
Usually, the A-mapping is used for non-methylated polymers, such as PS or PAM \cite{rossi_coarse-graining_2011, wang_coarse-grained_2012}, while the B-mapping is more common for representation of methylated polymers, such as PMMA\cite{uttarwar_study_2013}. Following this logic, we use the A-mapping for PNIPA.
For our Martini model, we reproduce bond-angle potentials as a harmonic angle potential:
\begin{equation}
\label{eq:3}
v_{\text{bond-angle}}(\theta_{ijk})=\tfrac{1}{2} k_{\theta}\big(\theta_{ijk}-\theta^{0}_{ijk}\big)^{2},
\end{equation}
here $\theta^{0}_{ijk}$ is the equilibrium angle, $k_{\theta}$ is the bond angle stiffness constant. We calculate $r_0$, $k_{\text{b}}$, $\theta^{0}_{ijk}$ and $k_{\theta}$ from atomistic simulations.

\subsection{Computational details}
\label{M&MC}
We use the latest DLPOLY $4.10$\cite{todorov_dl_poly_3:_2006} release for PNIPA modelling and DLPOLY Classic to calculate partitioning free energies by thermodynamic integration (TI). To parameterize the bond-length and the bond-angle interactions we apply the following procedure. First, we perform $\SI{30}{\nano \second}$ atomistic simulations of PNIPA of a length of 30 monomer units in TIP3P explicit water. We use the CHARMM-36 force field with the NVT ensemble. Temperature is maintained by a Nosé-Hoover themostat with $\SI{0.5}{\pico \second}$ coupling time. The reason we choose a stochastic thermostat rather than a velocity-rescale is that lately such thermostats have been critically reviewed\cite{braun_anomalous_2018}. We use the velocity Verlet integrator with a $\SI{2}{\femto\second}$ time step. The Lennard-Jones and short-range Coulomb potentials are shifted to zero at $\SI{1.2}{\nano\metre}$; the long-range electrostatic interactions are implemented using the particle mesh Ewald (PME) method. Having obtained atomistic trajectories, we calculate probability distributions for all possible CG bonds and valence angles and thus get a corresponding CG potential. Note that since we use the A-mapping scheme, we do not consider dihedral interaction in our CG model since they are already included in the bond-length and the bond-angle potentials. For CG calculations we use velocity Verlet integrator with with a $\SI{20}{\femto\second}$ time step. We use shifted straight cutoff Lennard-Jones potential which turns to zero at $\SI{1.2}{\nano\metre}$\cite{de_jong_martini_2016}; there are no electrostatic interactions since CG particles are uncharged.
\section{Results}
\subsection{Partitioning Free Energies}
\label{Res:A}

First of all, we have to parameterize PNIPA at different temperatures based on partitioning free energies between water and octanol-$1$. A mapping scheme of PNIPA is shown in Figure \ref{Plot:1}C). Following the previous works\cite{rossi_coarse-graining_2011,wang_coarse-grained_2012,banerjee_coarse-grained_2018} we use a SC$1$ bead to represent hydrocarbon groups of a backbone chain in the Martini $2.2$ and a TC$2$ bead in the Martini $3.0$ force-field. As for amide group, it seems reasonable to use a P- type bead; however, the level of polarity needs to be defined. Since we suppose that thermoresponsivity is connected with the ability of PNIPA to form hydrogen bonds with water, the polarity of the AM can be changed at different temperatures. As for isopropyl group, a C- type bead has to be used. \citeauthor{marrink_martini_2007} used a C$2$ bead to describe an IP group; in our case we take it as a starting point for parametrization.

To our knowledge, PNIPA partitioning free energies were not measured at wide temperature range. Thus we have to use data obtained by atomistic simulations, which is easily available in literature. In our case we would like to parameterize PNIPA at $\SI{280}{K}$ and at $\SI{330}{K}$ and we use the values, that is, at temperatures below and above LCST. Different force fields yield different partitioning free energies at these temperatures and to compare the results obtained in Martini we choose CHARMM force field as it describes well PNIPA partitioning at least at $\SI{300}{K}$\cite{kamath_thermodynamic_2013}.
\begin{table*}
\small
\caption{$\log{K_{OW}}$ for different types of beads representing AM (horizontal) and IP (vertical) groups of PNIPA for Martini $2.2$\label{Tab1} }
\begin{tabular*}{\textwidth}{@{\extracolsep{\fill}}c|cccc|ccc|ccc|ccc}
\hline
&\multicolumn{4}{c|}{$\SI{280}{K}$}&\multicolumn{3}{c|}{$\SI{300}{K}$}&\multicolumn{3}{c|}{$\SI{310}{K}$}&\multicolumn{3}{c}{$\SI{330}{K}$}\\
&P$2$&P$3$&P$4$&P$5$&P$3$&P$4$&P$5$&P$3$&P$4$&P$5$&P$1$&P$2$&P$3$\\
\hline
C$4$&$-$&$-$&$-$&$1.15$&$-$&$-$&$-$&$-$&$-$&$-$&$-$&$-$&$-$\\
C$5$&$0.96$&$-$&$0.54$&$0.35$&$0.73$&$0.56$&$0.46$&$1.14$&$1.04$&$0.6$&$2.77$&$2.65$&$1.65$\\
\hline
\end{tabular*}
\end{table*}
\begin{table*}
\small
\caption{$\log{K_{OW}}$ for different types of beads representing AM (horizontal) and IP (vertical) groups of PNIPA for Martini $3.0$\label{Tab2} }
\begin{tabular*}{\textwidth}{@{\extracolsep{\fill}}c|ccc|ccc|ccc|ccc}
\hline
&\multicolumn{3}{c|}{$\SI{280}{K}$}&\multicolumn{3}{c|}{$\SI{300}{K}$}&\multicolumn{3}{c|}{$\SI{310}{K}$}&\multicolumn{3}{c}{$\SI{330}{K}$}\\
&TP$3$&TP$4$&TP$5$&TP$3$&TP$4$&TP$5$&TP$3$&TP$4$&TP$5$&TP$3$&TP$4$&TP$5$\\
\hline
TC$3$&$1.47$&$0.98 $&$0.95 $&$2.24$&$1.66$&$1.63$&$2.34$&$1.93$&$1.9 $&$2.44$&$2.17$&$2.2 $\\
TC$4$&$1.13$&$0.59 $&$0.43 $&$1.43$&$1.07$&$0.99$&$1.55$&$1.15$&$1.15$&$1.75$&$1.31$&$1.27$\\
TC$5$&$0.61$&$-0.41$&$-0.24$&$0.99$&$0.7 $&$0.65$&$1.23$&$0.89$&$0.87$&$1.5 $&$1.11$&$1.11$\\
\hline
\end{tabular*}
\end{table*}
To determine exact types of AM and IP beads we calculate $\Delta\Delta G_{OW}$ for several possible combinations of beads. For that we find free energies of solvation in water and octanol-$1$ ($\Delta G_{W}$ and $\Delta G_{O}$, respectively) using thermodynamic integration. Free energy of partitioning is given by $\Delta\Delta G_{OW}=\Delta G_{W}-\Delta G_{O}$, so one can calculate $\log{K_{OW}}=\frac{\Delta\Delta G_{OW}}{k_{B}T}$. The results of these calculations for Martini $2.2$ are presented in Table \ref{Tab1}. We might notice that, considering the error of calculations being of $\SI{0.4}{\kilo\joule\per\mole}$, several configurations are possible at each temperature. Further on, we will test different possible representations of PNIPA. 


As for Martini $3.0$, the results of measuring $K_{OW}$ are given in Table \ref{Tab2}. Also, several representations are possible, although there are representations which correspond to all-atom data at every temperature, for instance TC$2$-TP$3$-TC$3$.

\subsection{Bonded interactions and representation of stereoregular effects}
\label{Res:B}
\begin{table*}
\small
\caption{Intramolecular potentials of PNIPA\label{Tab3} }
\begin{tabular*}{\textwidth}{@{\extracolsep{\fill}}cccccc}
\hline
{Valence bond}&{$r_{0}, \SI{}{\nano\metre} $}&{$k_{\text{b}}, \SI{}{\kilo\joule\per\mole\per\square\nano\metre}$} &{Valence angle}&{$\theta_{0}$}&{$k_{\theta}, \SI{}{\kilo\joule\per\mole}$}\\
\hline
{BB-AM}&$0.275$&$77500$&{BB-AM-BB}&$\frac{\pi}{3}$&$500$\\
{AM-IP}&$0.27$&$80000$&{AM-BB-AM (isotactic)}&$\frac{2 \pi}{3}$&$45$\\
&&&{AM-BB-AM (syndiotactic)}&$\pi$&$45$\\
&&&{BB-AM-IP}&$\frac{5 \pi}{6}$&$250$\\
\hline
\end{tabular*}
\end{table*}
To parameterize the bond-length and bond-angle interactions we apply the following procedure. First, we perform $\SI{30}{\nano\second}$ atomistic simulations of PNIPA of a length of $30$ monomer units. We use the CHARMM-$36$ force field with the NVT ensemble. Temperature is maintained by a Nosé-Hoover thermostat with $\SI{0.5}{\pico\second}$ coupling time. Then, we calculate probability distributions for all possible bonds and valence angles; these probabilities serve as target in the parametrization. We get the corresponding CG potentials by choosing the force constants and equilibrium lengths/angles to reproduce the width of the distributions and their centers, respectively. These values are given in Table \ref{Tab3}.

Usually stereoregular effects are not considered within the Martini force-field. However, such details are important for the description of PNIPA behaviour, thus it is crucial to represent them in the model. We make that by introducing different equilibrium angles AM-BB-AM for syndio- and isotactic chains.

\begin{figure*}
\centering
\includegraphics[width=13cm,trim=0 200 0 20,clip]{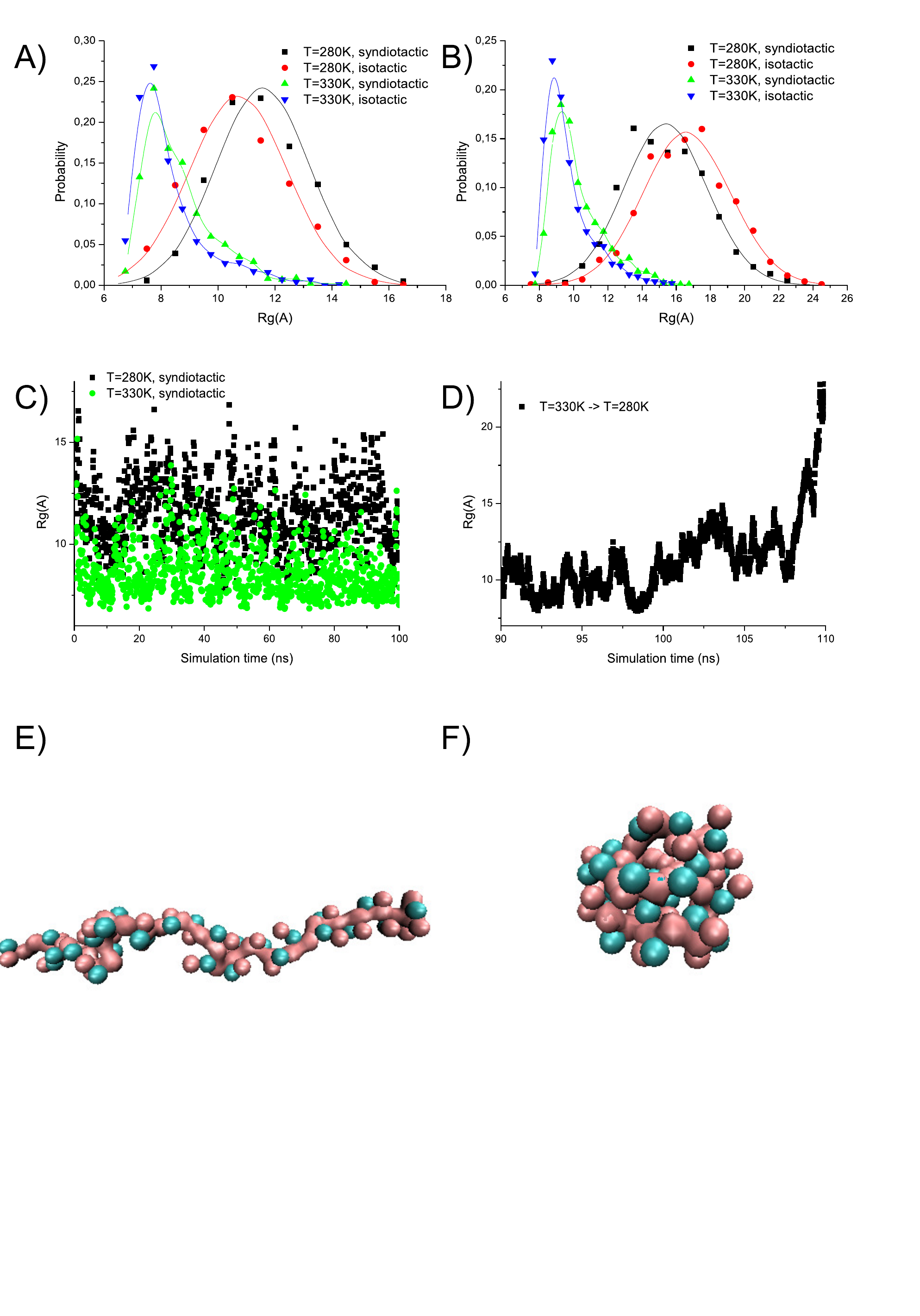}%
\caption{The PNIPA representation in Martini $2.2$. A-B) The probability distributions of the radius of gyration, $R_{g}$, for A) $N=20$; B) $N=30$ monomer units for syndiotactic/isotactic PNIPA at $\SI{280}{\K}$/$\SI{330}{\K}$; C) The radii of gyration versus time for isotactic PNIPA at $\SI{280}{\K}$/$\SI{330}{\K}$ for $N=20$ monomer units; D) The radius of gyration versus time for isotactic PNIPA when the temperature is decreased below LCST for $N=30$ monomer units; E-F) Snapshots of the isotactic PNIPA of $N=30$ monomer units at E)$\SI{280}{\kelvin}$ and F)$\SI{330}{\kelvin}$. \label{Plot:3}}%
\end{figure*}
\begin{figure*}
\centering
\includegraphics[width=13cm,trim=0 180 0 20,clip]{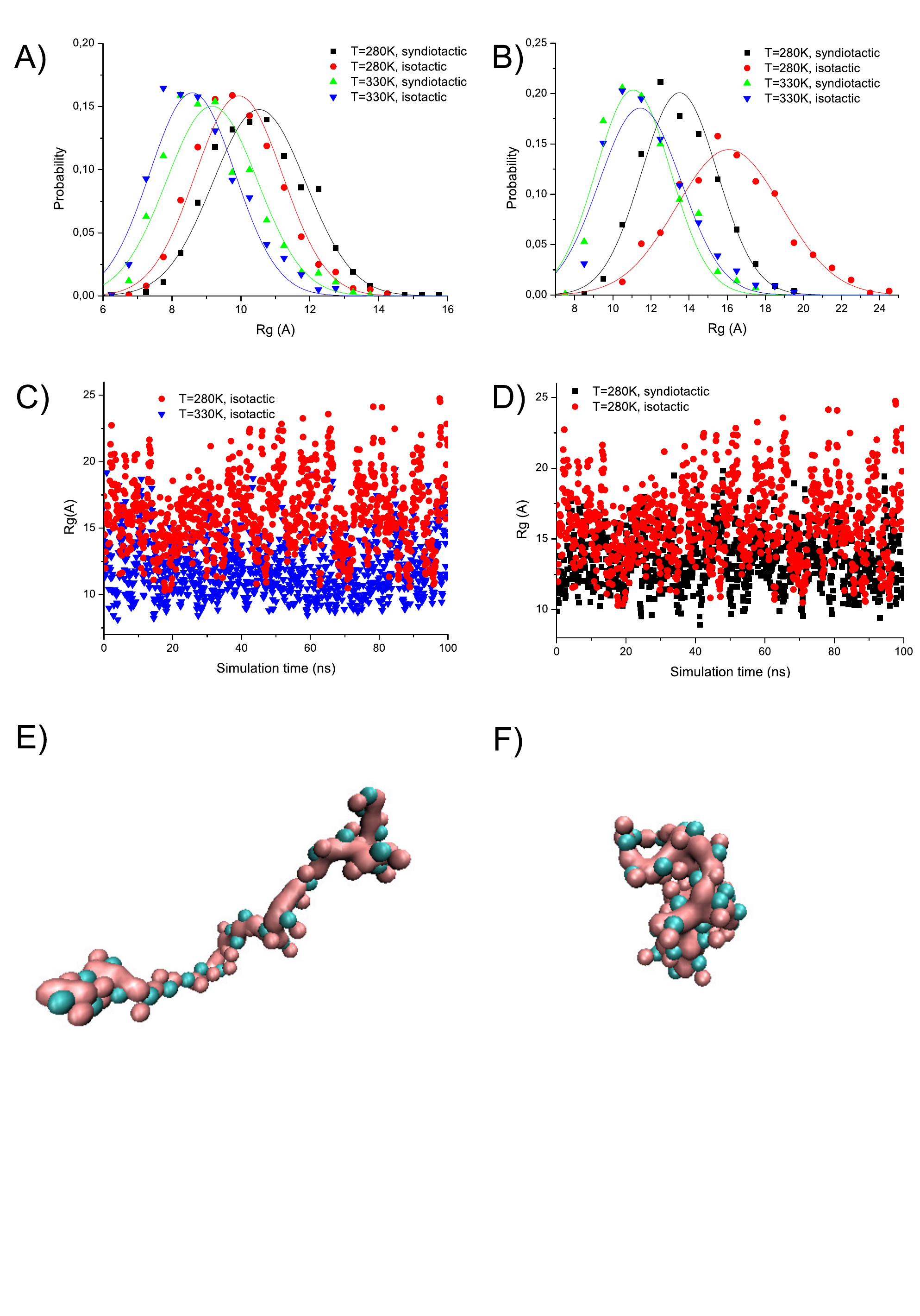}%
\caption{The PNIPA representation in Martini $3.0$. A-B) The probability distributions of the radius of gyration, $R_{g}$, for A) $N=20$; B) $N=30$ monomer units for syndiotactic/isotactic PNIPA at $\SI{280}{\K}$/$\SI{330}{\K}$; C) The radii of gyration versus time for isotactic PNIPA at $\SI{280}{\K}$ for $N=30$ monomer units; D) The radii of gyration versus time for syndio- and isotactic PNIPA at $\SI{280}{\K}$ for $N=30$ monomer units; E-F) Snapshots of the isotactic PNIPA of $N=30$ monomer units at E)$\SI{280}{\kelvin}$ and F)$\SI{330}{\kelvin}$. \label{Plot:4}}%
\end{figure*}

Below LCST both syndio- and isotactic polymers behave as flexible chains in both Martini $2.2$ (consult probability distribution: Figure \ref{Plot:3} A-B)) and Martini $3.0$ (see Figure \ref{Plot:4} A-B) and D)); above LCST they collapse into a globule. In Martini $2.2$ PNIPA tends to form denser globules than in Martini $3.0$; we suppose that this is due to the fact that Martini $3.0$ exploits the beads smaller in size. Anyway, our results for both Martini $2.2$ and Martini $3.0$ are in good correspondence with the ones previously reported for atomistic modelling\cite{tucker_study_2012}.

It is interesting to see whether a collapsed globule would become a coil when the temperature is lowered below LCST. Atomistic simulations fail to represent such transition, although it is reported to be observed experimentally. Our results show that PNIPA undergoes a globule-to-coil transition quite fast (see Figure \ref{Plot:3} D)), in approximately $\SI{10}{\nano\second}$. It might seem to be quite unexpected, since such simulation times are widely accessed by all-atom MD. However we are using a CG force-field wherein processes connected with some degrees of freedom occur much faster than in real life. In Martini, for instance, all hydrogen-bond forming abilities are accounted within Lennard-Jones potential, thus it is not that surprising that a decollapse occurs so rapidly.

At the same time, our findings show that a collapse in the Martini force-field occurs as fast as in all-atom simulations\cite{abbott_single_2015}. Also, our findings are in good accordance with the previously published coarse-grained models\cite{bejagam_machine-learning_2018,abbott_temperature-dependent_2015}.

\subsection{Choice of the representation at different temperatures}
\label{Res:C}

Partitioning free energies for both Martini $2.2$ and $3.0$ suggest several mappings at each temperature, as we have shown in the section \ref{Res:A}. For Martini $2.2$ we conduct $\SI{200}{\nano\second}$ simulations of PNIPA of lengths of $20$ and $30$ monomer units at each temperature using each possible representation (see Supporting Information). Having measured the radii of gyration of PNIPA, we note that none of the representations reproduces LCST; thus, we have to use different mappings for different temperatures. Below $\SI{280}{\kelvin}$ we choose C$1$-P$5$-C$5$, C$1$-P$4$-C$5$ at $\SI{300}{\kelvin}$ and C$1$-P$3$-C$5$ at temperatures above $\SI{310}{\kelvin}$.

For Martini $3.0$ we apply a similar procedure: we measure the radii of gyration of PNIPA of length of $30$ monomer units for each possible representation at different temperatures; unfortunately, no mapping reproduces LCST. Thus, the model has to be temperature-dependent and we choose TC$2$-TP$5$-TC$5$ representation at temperatures below $\SI{280}{\kelvin}$, TC$2$-TP$4$-TC$5$ at $\SI{300}{\kelvin}$ and TC$2$-TP$3$-TC$5$ at temperatures above $\SI{310}{\kelvin}$.

\subsection{Choice of the particle size and bonded interactions}
\label{Res:D}
\begin{figure}
\centering
\includegraphics{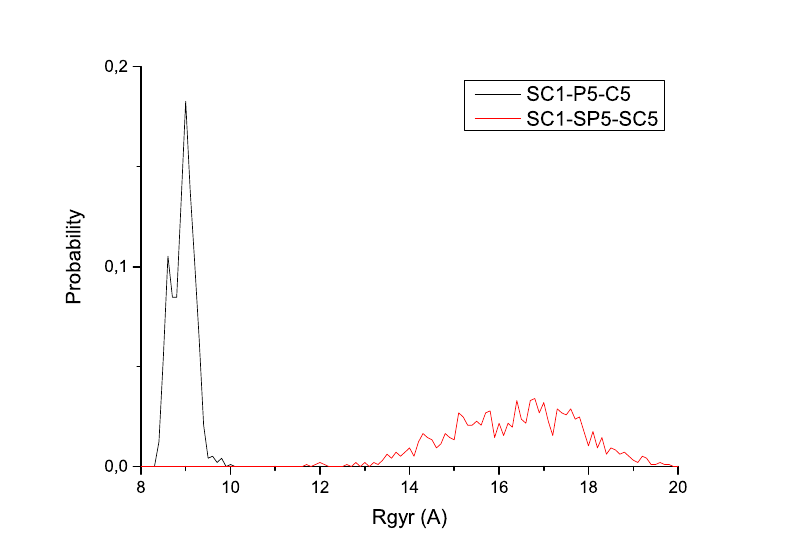}%
\caption{The probability distributions of the radius of gyration, $R_{g}$ , at $\SI{280}{\K}$ for isotactic PNIPA of $N=30$ monomer units for the SC$1$-P$5$-C$5$ and SC$1$-SP$5$-SC$5$ representations \label{Plot:5}}%
\end{figure}

As mentioned before (see Section \ref{M&MA}), in the Martini force-field coarse-grained particles are distinguished by their size. We apply $3$-to-$1$ mapping for both AM and IP groups; for such mapping both N-beads \cite{marrink_martini_2007} and S-beads\cite{nawaz_coarse-graining_2014,banerjee_coarse-grained_2018,lee_coarse-grained_2009} were used. Size of the particle cannot be defined while measuring partitioning free energy between water and octanol-$1$ since Martini $2.2$ solvents are represented by N-beads. Thus we performed separate simulations of PNIPA in which atoms of side groups were represented by N-beads and S-beads, the probability distributions are given in Figure \ref{Plot:5}. Simulations show that if we use N-beads, PNIPA collapses at $\SI{280}{\K}$.

When S-beads are concerned, PNIPA remains in the coiled state. Thus S-beads make better representation of PNIPA. This might be due to the fact that C$5$ particles representing an IP group screen highly polar P$5$ particles from the interaction with water while SC$5$ particles let the contacts between AM and water particles occur.

\section{Discussion}

To the best of our knowledge, our approach is one of the first reported implementations of the latest Martini $3.0$\cite{souza_proteinligand_2020}. We note that this version is more accurate and flexible than the old one, since it does not limit the usage of T- and S-beads for rings and introduces new interaction levels, which results in smoother transition between beads.

Another interesting finding is that we managed to capture the difference between iso- and syndiotactic polymers. Usually, only atactic chains are concerned for parametrization of polymers \cite{rossi_coarse-graining_2011, banerjee_coarse-grained_2018}; therefore either the difference between meso- and racemic diads is integrated out by applying a harmonic angle/cosine angle potential\cite{rossi_coarse-graining_2011, wang_coarse-grained_2012} or complicated multi-centered potentials have to be applied to correctly capture the above-mentioned difference\cite{abbott_temperature-dependent_2015, banerjee_coarse-grained_2018}. The latter limits the timestep used in simulation to approximately $\SI{10}{\femto \second}$. In our case we set simple, but different angle bond potentials for meso- and racemic diads; this seems to be reasonable for the considered simulations since a racemic diad is unable to transform into a meso one in real polymer chains. Thus we can use larger timesteps of $\SI{20}{\femto \second}$ and our results remain in accordance with the data obtained from atomistic MD. To our knowledge, our implementation is one of the first to demonstrate the difference between iso- and syndiotactic polymers within a coarse-grained model.

It is also important to notice that within our model PNIPA "senses" cooling below the LCST, that is it transforms from the globular state to the coiled state. Although PNIPA has been repeatedly reported to undergo such a transition in experiments, all atom MD was not able to detect it even on long trajectories (over $\SI{100}{\nano \second}$) \cite{abbott_single_2015, tucker_study_2012}. Seemingly, the lack of explicit hydrogen bonds in the CG Martini force-field (all hydrogen-bonding capabilities are included into Lennard-Jones potentials) facilitates this process in our model.

Comparing our model with the other developed CG models, we think that it can be used much wider and boarder than those previously introduced. Our results are in agreement with the SDK model by \citeauthor{abbott_temperature-dependent_2015}, which is also temperature-dependent. Their model produces similar $R_{g}$ distributions. However, the Martini force-field is much more widespread than SDK, and thus our model is thought to be more accessible for simulations. As for the model by \citeauthor{bejagam_machine-learning_2018}, its main downfall is its specificity; it correctly reproduces PNIPA's behaviour and properties, but it might need additional parametrization should one need to study the PNIPA interaction with other compounds; In this sense the Martini force-field is much more applicable and flexible. 

Also we would like to compare our model with that recently introduced by \citeauthor{perez-ramirez_coil--globule_2020}. As we have previously mentioned, our main concern is that their model, correct as it is, does not comply with Martini since the values used in modelling contradict the standard ones. Thus, technically speaking, their model is an excellent mesoscopic coarse-grained model capturing LCST of PNIPA at $\SI{302.1}{\kelvin}$, but it cannot be referred to Martini. We suppose that the implemented modifications have successfully led to the correct LCST, but one have to choose between the fulfilment with Martini parameters (which leads to compatibility with all other parametrized compounds) or correspondence to all-atom data.

The choice between different bead sizes is another interesting issue. In the work by \citeauthor{marrink_martini_2007} mapping of almost any chemical group was performed by N-beads, while S-beads were reserved for ring structures exclusively. Later, S-beads started being exploited more often\cite{lee_coarse-grained_2009,taddese_effect_2017,banerjee_coarse-grained_2018}: for instance, following \citeauthor{nawaz_coarse-graining_2014} even four atoms of poly(propylene oxide) monomer unit are mapped onto one S-bead. However, the difference between S- and N-beads cannot be captured by partitioning coefficients between water and octanol-$1$ due to the fact that S-beads interact with N- ones as N-beads while both water and octanol-$1$ are represented by N-beads (P$4$ and P$1$-C$1$, respectively). It is worth noting that two more widespread solvents were considered by \citeauthor{marrink_martini_2007}: benzene, represented by three SP$4$ particles, and hexane, represented by two C$1$ particles. But following the logic behind latter works on the Martini force-field, in which three atoms are mapped onto S-bead, both octanol-$1$ and hexane as well might be represented by three and two S-beads respectively. Actually \citeauthor{taddese_effect_2017}, parameterizing poly(ethylene oxide), have considered both N- and S- hexane. Such ambiguity leads to confusion, thus intuitively we intend to distinguish between S-beads representing two or three atoms. Unfortunately, such a challenge is well out of the scope of this work, and will be a subject of our further research.

\section{Conclusion}
Our model takes into account polymer's stereoregularity (i.e. tacticity) and correctly represents the difference between isotactic and syndiotactic PNIPA; the results are in good accordance with atomistic simulations. Moreover, the developed CG model represents globule-to-coil transition, which is still unaccessible by all-atom MD. The proposed approach might be advantageously exploited to study other thermoresponsive and/or stereo(ir)regular polymers; it allows to address big time- and lenghtscales.

\section*{Conflicts of interest}
There are no conflicts to declare.

\section*{Acknowledgements}
This work was partly supported within framework of the state task for Federal Research Center for Chemical Physics RAS $\#$FFZE-$2019$-$0016$.

\bibliographystyle{unsrtnat}

\end{document}